\documentclass[twocolumn]{revtex4}
\usepackage{epsfig}
\usepackage{amssymb}
\usepackage{amsmath}
\usepackage{amsfonts}
\usepackage{graphicx}
\usepackage{mathrsfs}
\usepackage[dvips]{color}
\usepackage{multirow}

\newcommand{\fn}{\mathfrak{n}}

\newcommand{\fz}{\mathfrak{z}}

\newcommand{\fK}{\mathfrak{K}}

\newcommand{\cR}{\mathcal{R}}

\newcommand{\be}{\begin{equation}}
\newcommand{\ee}{\end{equation}}
\newcommand{\bea}{\begin{eqnarray}}
\newcommand{\eea}{\end{eqnarray}}
\newcommand{\nn}{\nonumber}

\newcommand{\ed}{\end{document}}

\newcommand{\rx}{{\rm x}}
\newcommand{\rg}{{\rm g}}
\newcommand{\ry}{{\rm y}}

\newcommand{\bi}{\begin{itemize}}
\newcommand{\ei}{\end{itemize}}

\newcommand{\bce}{\begin{center}}
\newcommand{\ece}{\end{center}}

\newcommand{\RE}{\,{\rm Re}}
\newcommand{\IM}{\,{\rm Im}}

\newcommand{\sE}{\mathscr{E}}

\begin{document}
\title{Nonlinear Spectral Singularities and Lasing Threshold Condition}
\author{Ali~Mostafazadeh}
\address{Department of Mathematics, Ko\c{c}
University,\\ Sar{\i}yer 34450, Istanbul, Turkey\\
amostafazadeh@ku.edu.tr}

\begin{abstract}
A spectral singularity is a mathematical notion with an intriguing physical realization in terms of certain zero-width resonances. In optics it manifests as lasing at the threshold gain. We explore the application of their recently-developed nonlinear generalization in the study of the effect of a nonlinearity on the lasing threshold condition for an infinite planar slab of gain medium. In particular, for a Kerr nonlinearity, we derive an explicit expression for the intensity of the emitted waves from the slab and discuss the implications of our results for the time-reversed system that acts as a coherent perfect absorber. \\[3pt]
\hspace{6.2cm}{Pacs numbers: 42.25.Bs, 42.65.-k, 24.30.Gd, 03.65.Nk}
\end{abstract}

\maketitle

\section{Introduction}
One of the most interesting properties of complex scattering potentials \cite{muga} is that unlike their real analogs they can admit scattering states satisfying Siegert outgoing boundary conditions \cite{siegert}. Because these states have a real and  positive energy, they behave exactly like a zero-width resonance. This observation is originally made in \cite{prl-2009} in an attempt to understand the physical meaning of the mathematical notion of a spectral singularity \cite{naimark}. It has motivated the study of physical aspects and applications of spectral singularities particularly in optics \cite{prl-2009,pra-2009,optical-ss,CPA,pra-2011a,pra-2011b,pra-2012}.

A remarkable outcome of this study is that for an optical system consisting of an infinite planar slab of homogeneous gain medium the well-known lasing threshold condition follows from the equation ensuring the generation of a spectral singularity \cite{pra-2011a}. Furthermore, if we consider the time-reversed setup, where the slab consists of a lossy medium with the loss coefficient having the same value as the gain coefficient required for a spectral singularity, it behaves as a coherent perfect absorber \cite{CPA}.

A more recent development is a nonlinear generalization of the notion of a spectral singularity \cite{p110}. A characteristic feature of nonlinear spectral singularities is that they are intensity-dependent. This means that they correspond to the emission of waves with a particular wavelength-amplitude profile. Ref.~\cite{p110} offers details of this phenomenon for a delta-function potential with a complex coupling constant \cite{jpa-2006}.

In order to improve our understanding of the properties and possible applications of nonlinear spectral singularities, in the present article, we study their behavior for an infinite planar slab of gain material subject to a weak nonlinearity. In particular, we give a detailed analytic description of the effects of a Kerr nonlinearity.

Consider a linearly polarized electromagnetic wave that is aligned along the $x$-axis in some
cartesian coordinate system and has an electric field of the form
$\vec E= e^{-i\omega t}\sE(z)\hat e_x$, where $\omega$ is the angular frequency of the wave, $\sE$ is a complex-valued function, and $\hat e_x$ is the unit vector along the positive $x$-axis. Suppose that this wave interacts with an infinite planar slab of gain material of thickness $a$ that is aligned in the $x$-$y$ plane and located between the planes $z=0$ and $z=a$. Then it is well-known that the Helmholtz equation for this system reduces to the Schor\"odinger equation,
    \be
    -\sE''(z)+V(z)\sE(z)=k^2\sE(z),
    \label{sch-eq}
    \ee
for the energy-dependent complex barrier potential \cite{pra-2011a}
    \be
    V(z):=k^2(1-\fn^2)\chi_{a}(z),
    \label{V=}
    \ee
where $k:=\omega/c$ is the wavenumber, $\fn$ is the complex refractive index of the slab, and
    \be
    \chi_{a}(z):=\left\{\begin{aligned}
    &1&&{\rm for}~~0\leq z\leq a,\\
    & 0 &&~~{\rm otherwise.}
    \end{aligned}\right.
    \ee

When a nonlinearity is present, $\fn^2$ is replaced by $\fn^2+\rg f(|\sE(z)|)$ and (\ref{sch-eq}) takes the form
    \be
    -\sE''(z)+V(z)\sE(z)-\rg\, k^2\,\chi_{a}(z)f(|\sE(z)|)\sE(z)=k^2\sE(z),
    \label{nl-sch-eq-1}
    \ee
where $\rg$ is a real constant and $f$ is a real-valued function \cite{eberly}. In terms of the scaled variables,
    \be
    \begin{aligned}
    &\rx:=z/a, &&\psi(x):=\sE(a\,\rx), &&\fK:=a k,\\
    &\gamma:=-\fK^2\rg,&&\fz:=\fK^2(1-\fn^2), &&v(\rx):=\fz\,\chi_1(\rx),\\
    \end{aligned}
    \label{scale}
    \ee
we can write (\ref{nl-sch-eq-1}) as
    \be
    -\psi''(\rx)+v(\rx)\psi(\rx)+\gamma\chi_{1}(\rx)f(|\psi(\rx)|)\psi(\rx)=\fK^2\psi(\rx).
    \label{nl-sch-eq}
    \ee
For a Kerr nonlinearity, where
    \be
    f(|\psi|)=|\psi|^2,
    \label{Kerr}
    \ee
(\ref{nl-sch-eq}) is known as the Gross-Pitaevskii equation and has applications in the description of the Bose-Einstein condensates \cite{rwk}. For  real values of $\fz$ it admits exact analytic solutions in terms of the Jacobi elliptic functions \cite{CCR}.

\section{Nonlinear Spectral Singularities}

In Ref.~\cite{p110} we introduce nonlinear spectral singularities associated with (\ref{nl-sch-eq}). In short, they correspond to the real and positive values of $\fK^2$ for which (\ref{nl-sch-eq}) admits a nonzero solution fulfilling the outgoing boundary conditions:
	\be
	\psi'(0)+i\fK\,\psi (0)=0,~~~~
	\psi'(1)-i\fK\,\psi_k(1)=0.
	\label{b-condi}
	\ee
In order to see the physical implications of these conditions, we use the fact that for $\rx\notin[0,1]$ the general solution of (\ref{nl-sch-eq}) is a linear combination of the plane waves $e^{\pm i\fK\rx}$. This allows us to envisage scattering solutions of  (\ref{nl-sch-eq}) corresponding to incident waves from the left ($\rx<0$) and the right ($\rx>1$). The former has the form \cite{p110}:
    \be
    \psi^l(\rx)=
    \left\{\begin{array}{ccc}
	N_-  e^{-i\fK\rx}+\tilde N_-  e^{i\fK\rx}
    &{\rm for}& \rx<0,\\[3pt]
    \zeta(\rx)&{\rm for}& 0\leq\rx\leq 1,\\[3pt]
    N_+  e^{i\fK\rx} &{\rm for}& \rx> 1,
    \end{array}\right.
    \label{jp}
    \ee
where $\zeta$ is a solution of (\ref{nl-sch-eq}) that satisfies
	\begin{align}
	&\zeta(1)=N_+ e^{i\fK}, &&\zeta'(1)=i\fK N_+ e^{i\fK},
	\label{zeta}
	\end{align}
$N_+$ is a complex constant that gives the amplitude of the transmitted wave, and
	\begin{align}
	&N_-:=\frac{iG_-(\fK)}{2\fK},\quad\quad\quad\quad
	\tilde N_-:=-\frac{iG_+(\fK)}{2\fK},
	\label{Ns}\\[3pt]
	&G_{\pm}(\fK):=\zeta'(0)\pm i \fK\,\zeta(0).
	\label{Gs}
	\end{align}

We can use (\ref{jp}) and (\ref{Ns}) to derive the following expressions for the reflection and transmission coefficients from the left, respectively \cite{p110}.
    \begin{align}
    &R^l=-\frac{G_-(\fK)}{G_+(\fK)}, && T^l=\frac{2ik N_+}{G_+(\fK)}.
	\label{R-T}
    \end{align}
In view of (\ref{zeta}) the solution~(\ref{jp}) satisfies the second equation in (\ref{b-condi}). The first of these equations, which is equivalent to $G_+(\fK)=0$, implies that both the reflection and transmission coefficients diverge. This is a characteristic property of a resonance state. Notice however that the energy $\fK^2$ does not have an imaginary part.  Therefore (\ref{jp}) is a scattering solution of (\ref{nl-sch-eq}) which behaves like a resonance with a zero width \cite{prl-2009}.

Now, suppose that we tune the parameters of the system so that it supports a spectral singularity with  wavenumber $k$ and amplitude parameter $N_+$. Then any left-incident plane wave with wavenumber and amplitude, respectively, in a close vicinity of $k$ and $\tilde N_-$ is amplified and emitted from both sides of the slab. In the absence of the nonlinearity, this phenomenon is amplitude-independent and the system amplifies the background noise and serves as a laser that functions at the very threshold gain \cite{pra-2011a}. The presence of the nonlinearity modifies the lasing threshold condition by introducing in it a particular amplitude-dependence. Our main objective is to derive an explicit expression for this modified lasing threshold condition.

\section{Lasing Threshold Condition}

Determination of nonlinear spectral singularities requires the solution of the boundary-value problem defined by (\ref{nl-sch-eq}) and (\ref{b-condi}). This is equivalent to finding the function $\zeta$, which satisfies (\ref{nl-sch-eq}) and (\ref{zeta}), and demanding that
    \be
    G_+(\fK)=0.
    \label{G=0}
    \ee
For $\rx\in[0,1]$, we can express (\ref{nl-sch-eq}) in the form of the integral equation
    \be
    \psi(\rx)=\psi_0(\rx)+\gamma\int_{\rx_0}^\rx G(\rx-\ry)f(|\psi(\ry)|)\psi(\ry)d\ry,
    \label{int-eq}
    \ee
where $\psi_0$ is the general solution of the linear equation
    \be
    \psi''+\fn^2\fK^2\psi=0,
    \label{homog}
    \ee
in $[0,1]$, $\rx_0\in[0,1]$ is arbitrary, and $G$ is the Green's function for this equation, i.e.,
    \be
    G(u):=\frac{\sin(\fn\fK\,u)}{\fn\fK}.
    \ee
Repeated use of (\ref{int-eq}) in its right-hand side yields a perturbative series expansion for $\psi$ with $\gamma$ playing the role of the perturbation parameter. In particular, we find the following first-order perturbative expression for $\zeta$.
    \be
    \zeta(\rx)\approx \zeta_0(\rx)+\gamma\,\zeta_1(\rx),
    \label{zeta=}
    \ee
where $\approx$ means that we neglect quadratic and higher order terms in powers of $\gamma$, $\zeta_0$ is the solution of (\ref{homog}) satisfying (\ref{zeta}), and
    \be
    \zeta_1(\rx):=\int_{1}^\rx G(\rx-\ry)f(|\zeta_0(\ry)|)\,\zeta_0(\ry)d\ry.
    \label{z1}
    \ee
It is easy to show that
    \be
    \zeta_0(\rx)=\frac{N_+ e^{i\fK}}{2\fn}\left[(\fn+1)e^{i\fn\fK(\rx-1)}+
    (\fn-1)e^{-i\fn\fK(\rx-1)}\right].
    \label{z0}
    \ee

Next, we insert (\ref{zeta=}) in (\ref{Gs}) to obtain
    \be
    G_\pm(\fK) \approx G_\pm^{(0)}(\fn,\fK)+\gamma G_\pm^{(1)}(\fn,\fK),
    \label{Gs-1}
    \ee
where
    \be
    G_\pm^{(j)}(\fn,\fK):=\zeta'_j(0)\pm i\fK \zeta_j(0)~~{\rm for}~~j=0,1.
    \label{Gs-2}
    \ee
The computation of $G_\pm^{(0)}(\fK)$ is straightforward. It gives
    \bea
    G_-^{(0)}(\fn,\fK)&=&\frac{N_+ e^{i\fK}\fK(\fn^2-1)\sin(\fn\fK)}{\fn},
    \label{G-m}\\
    G_+^{(0)}(\fn,\fK)&=&\left[\frac{i N_+ e^{i(\fn+1)\fK}(\fn+1)^2}{2\fn}\right]L(\fn,\fK),
    \label{G-p}
    \eea
where
    \be
    L(\fn,\fK):= e^{-2i\fn\fK}-\left(\frac{\fn-1}{\fn+1}\right)^2.
    \label{F=}
    \ee
According to (\ref{G-p}) and (\ref{F=}), in the absence of nonlinearity, $\gamma=0$, and we find a spectral singularity provided that the right-hand side of (\ref{F=}) vanishes. This is in complete agreement with the results of \cite{pra-2011a,pra-2011b,pra-2012}.

Let $\fn_0$ and $\fK_0$ be such that $L(\fn_0,\fK_0)=0$, i.e., they correspond to a linear spectral singularity, and $\eta_0$ and $\kappa_0$ be respectively the real and imaginary parts of $\fn_0$, so that $\fn_0:=\eta_0+i\kappa_0$. Then in view of (\ref{F=}) and the fact that for typical gain media,
    \be
    \eta_0-1\gg-\kappa_0>0,
    \label{real}
    \ee
we have
    \be
    e^{-i\fn_0\fK_0}=\frac{\fn_0-1}{\fn_0+1}.
    \label{eq1}
    \ee
Recalling that the gain coefficient $g_0$ necessary for the emergence of the spectral singularity is related to $\kappa_0$ according to $g_0=-2\fK_0\,\kappa_0/a$, and taking the modulus-square of both sides of (\ref{eq1}), we obtain
    \be
    g_0=\frac{1}{2a}\ln\frac{1}{|\cR|^2},
    \label{LTC0}
    \ee
where $\cR:=(\fn_0-1)/(\fn_0+1)$. In view of (\ref{real}), $\cR\approx(\eta_0-1)/(\eta_0+1)$. Therefore, it gives the reflexivity of the slab, and (\ref{LTC0}) coincides with the standard expression for the lasing threshold condition \cite{pra-2011a}, 
    \be
    g_0\approx\frac{1}{a}\ln\left(\frac{\eta_0+1}{\eta_0-1}\right).
    \label{LTC}
    \ee

In order to obtain nonlinear corrections to (\ref{LTC}), we solve (\ref{G=0}) perturbatively. Let $\fn_1$ and $\fK_1$ be such that, up to linear terms in $\gamma$, Eq.~(\ref{G=0}) holds for
    \begin{align}
    &\fn=\fn_0+\gamma\fn_1, &&\fK=\fK_0+\gamma\fK_1.
    \label{n-K=}
    \end{align}
Inserting (\ref{n-K=}) in (\ref{Gs-1}), expanding the resulting expression for $G_+$ in powers of $\gamma$, and equating the coefficient of $\gamma$ to zero give
    \be
    \partial_{\fn_0}G_+^{(0)}(\fn_0,\fK_0)\fn_1+
    \partial_{\fK_0}G_+^{(0)}(\fn_0,\fK_0)\fK_1+
    G_+^{(1)}(\fn_0,\fK_0)=0.
    \label{SS-eq}
    \ee
Eq.~(\ref{eq1}) simplifies the calculation of $\partial_{\fn_0}G_+^{(0)}(\fn_0,\fK_0)$ and $\partial_{\fK_0}G_+^{(0)}(\fn_0,\fK_0)$ enormously. The result is
    \bea
    \partial_{\fn_0}G_+^{(0)}(\fn_0,\fK_0)&=&
    \frac{N_+e^{i\fK_0}\fK_0\left[(\fn_0^2-1)\fK_0-2i\right]}{\fn_0},
    \label{eq11}\\
    \partial_{\fK_0}G_+^{(0)}(\fn_0,\fK_0)&=&N_+ e^{i\fK_0}\fK_0(\fn_0^2-1).
    \label{eq12}
    \eea
The calculation of $G_+^{(1)}(\fn_0,\fK_0)$ is more involved. However, it turns out that we can use (\ref{z1}), (\ref{Gs-2}), and (\ref{eq1}) to write it in the form:
    \be
    G_+^{(1)}(\fn_0,\fK_0)=
    -N_+ e^{i\fK_0}c_{\fn_0}^2\int_0^1 f\left(|N_+c_{\fn_0} h(x)|\right)h(x)^2 dx,
    \label{Gp=1}
    \ee
where
    \begin{align*}
    &c_{\fn_0}:=\frac{\fn_0+1}{2\fn_0}, && h(x):=e^{i\fn_0\fK_0(x-1)}+e^{-i\fn_0\fK_0x}.
    \end{align*}
For example, consider a Kerr nonlinearity (\ref{Kerr}). Then we can easily perform the integral in (\ref{Gp=1}) and use (\ref{eq1}) to express $G_+^{(1)}(\fn_0,\fK_0)$ as
    \be
    G_+^{(1)}(\fn_0,\fK_0)=\frac{8 i |N_+|^2N_+ e^{i\fK_0}(4\fn_0^2-\fn_0^{*2}-3) }{\fK_0(9\fn_0^4+\fn_0^{*4}-10|\fn_0|^4)}.
    \label{Gp=2}
    \ee

We can determine the nonlinear corrections to the location of the spectral singularities by substituting (\ref{eq11}), (\ref{eq12}), and (\ref{Gp=2}) in (\ref{SS-eq}) and solving the resulting equation for any two of $\RE(\fn_1)$, $\IM(\fn_1)$, and $\fK_1$ in terms of the third.

In order to gain a better understanding of the consequences of (\ref{SS-eq}), we expand the right-hand sides of (\ref{eq11}), (\ref{eq12}), and (\ref{Gp=2}) in powers of $\kappa_0$ and keep the leading order term. This gives
    \bea
    \partial_{\fn_0}G_+^{(0)}(\fn_0,\fK_0)&\approx&
    \frac{N_+e^{i\fK_0}\fK_0\left[(\eta_0^2-1)\fK_0-2i\right]}{\eta_0},~~~
    \label{eq11a}\\
    \partial_{\fK_0}G_+^{(0)}(\fn_0,\fK_0)&\approx&N_+ e^{i\fK_0}\fK_0(\eta_0^2-1),
    \label{eq12a}\\
     G_+^{(1)}(\fn_0,\fK_0)&\approx&\frac{3 |N_+|^2N_+ e^{i\fK_0}(\eta_0^2-1)}{4\eta_0^3\fK_0\,\kappa_0}.
    \label{Gp=2a}
    \eea

The gain coefficient $g$ that is required to support a nonlinear spectral singularity is given by
    \be
    g=-\frac{2\,\fK\,\kappa}{a}\approx g_0\left[1+\gamma\left(\frac{\kappa_1}{\kappa_0}+\frac{\fK_1}{\fK_0}\right)\right],
    \label{g-3}
    \ee
where $\kappa:=\IM(\fn)$ and $\kappa_1:=\IM(\fn_1)$. Suppose that we are interested in a nonlinear spectral singularity corresponding to a Kerr nonlinearity that has the same wavelength as its linear analog. Then $\fK_1=0$ and we can use (\ref{SS-eq}) and (\ref{eq11a}) -- (\ref{g-3}) to determine the modified lasing threshold condition:
    \bea
    g&\approx& g_0\left[1-\frac{6(\eta_0^2-1)\gamma |N_+|^2}{\eta_0^2\left[(\eta_0^2-1)^2\fK_0^2+4\right]
    \ln^2\!\!\left(\frac{\eta_0+1}{\eta_0-1}\right)}\right]
    \nn\\
    &\approx& g_0\left[1+\frac{6\,\rg |N_+|^2}{\eta_0^2 (\eta_0^2-1)
    \ln^2\!\!\left(\frac{\eta_0+1}{\eta_0-1}\right)}\right].
    \label{g=2}
    \eea
Here in the last relation we have switched to the unscaled nonlinearity parameter $\rg$ and use the fact that for typical optical setups $(\eta_0^2-1)^2/\fK_0^{2}\approx 0$.

Eq.~(\ref{g=2}) is our main result. It implies that for a positive Kerr nonlinearity parameter $\rg$, which is usually the case \cite{eberly}, the presence of the nonlinearity increases the necessary gain for producing a spectral singularity. More importantly it shows that the slab acts as an amplifier for a left-incident plane waves only if $g$ exceeds $g_0$. In this case it emits an amplified transmitted plane wave whose intensity is given by
    \be
    \frac{1}{2}|N_+|^2\approx \left(\frac{g-g_0}{12\,\rg\,g_0}\right)\!\!\left[\eta_0^2 (\eta_0^2-1)
    \ln^2\!\!\left(\frac{\eta_0+1}{\eta_0-1}\right)\right].
    \label{intensity}
    \ee
Because this relation follows from our first-order perturbative analysis, it applies for situations where $\rg|N_+|^2\ll 1$. Typically $\rg<10^{-13}\,{\rm cm}^2/{\rm W}$ which for a value of $|N_+|^2$  as large as $1~{\rm GW}/{\rm cm}^2$ gives $\rg|N_+|^2<10^{-4}$. This shows that (\ref{intensity}) is quite reliable.

We can view (\ref{intensity}) as a mathematical justification for the well-known fact that lasers operate when the gain coefficient is larger than $g_0$. It is important to notice that in the linear case the spectral singularity disappears as the value of the gain coefficient exceeds $g_0$ and the slab stops functioning as a laser. This is not the case when we take into account the effect of a Kerr nonlinearity with $\rg>0$.

We have performed a first-order perturbative calculation of $G_-(\fK)$ and used the result together with (\ref{Ns}) to compute the amplitude parameter $N_-$ for the emitted wave from the left-hand side of the slab. This calculation shows that whenever we arrange the parameters of the system to generate a nonlinear spectral singularity, the contribution of the terms of the order  $\gamma$ to $N_-$ cancel one another and we find $N_-\approx N_+ e^{i\fK}$. Therefore similarly to the linear case, the wave emitted from the left-hand side of the slab has the same amplitude and phase as the one emitted from its right-hand side. For the time-reversed system, this implies that a lossy slab serves as a coherent perfect absorber provided that the loss factor $\alpha$ has a numerical value that is larger than $g_0$ and the intensity $|N_+|^2/2$ of the incident coherent waves be given by (\ref{intensity}) with $g$ replaced with $\alpha$.

\section{Concluding Remarks}

An interesting outcome of the search for physical implications of spectral singularities is a mathematical derivation of the lasing threshold condition. This is quite different from the standard derivation of this condition which is essentially based on the physical principle of conservation of energy \cite{silfvast}. In the present article we have used the newly developed concept of a nonlinear spectral singularity to account for the contribution of nonlinearities to the lasing threshold condition.

Our method is quite general and can be applied to almost all nonlinearities of physical interest. For a Kerr nonlinearity it provides a simple explanation for the well-known fact that our slab model does not function as a laser at the threshold gain $g_0$. It starts emitting radiation only when the gain coefficient $g$ exceeds $g_0$. In this case, the intensity of the emitted radiation turns out to be proportional to $g-g_0$. \vspace{6pt}\\

\noindent\emph{Acknowledgments}.---  We wish to thank Aref Mostafazadeh, Alphan Sennaro\u{g}lu, and Ali Serpeng\"{u}zel for useful discussions. This work has been supported by  the Scientific and Technological Research Council of Turkey (T\"UB\.{I}TAK) in the framework of the project no: 110T611, and by the Turkish Academy of Sciences (T\"UBA).

\ed